# The origin of indistinguishability of quantum features in an interferometric system

Byoung S. Ham

School of Electrical Engineering and Computer Science, Gwangju Institute of Science and Technology

123 Chumdangwagi-ro, Buk-gu, Gwangju 61005, South Korea

(Submitted on Oct. 27, 2021; bham@gist.ac.kr)

**Abstract**
One of the most fundamental quantum features is the two-photon intensity correlation on a beam splitter (BS), resulting in photon bunching into either output port. According to the conventional understanding of quantum mechanics, the origin of photon bunching on a BS is in the indistinguishable characteristics between coincident photons, resulting in destructive quantum interference even without a clear definition of phase information of the paired photons. Here, a completely different approach is presented for the same quantum feature, where a strong mutual phase dependency is the essential requirement for the nonclassical feature of photon bunching. This definite phase relationship is now understood as the origin of this quantum feature resulting from the phase-basis superposition of the photon-BS system, resulting in photon indistinguishability. On behalf of a coherent model with a single input-port BS system, an extended scheme of a Mach-Zehnder interferometer is additionally analyzed for the validity of forbidden phase-basis superposition in the quantum feature.

**Introduction**
Two-photon intensity correlation has been the key concept of quantum mechanics since the seminal paper by Hanbury Brown and Twiss in 1957 [1]. Regarding photon entanglement, the Hong-Ou-Mandel (HOM) experiment demonstrated the nonclassical feature of anticorrelation via two-photon intensity correlation on a beam splitter (BS) in 1987 [2]. The entangled photon pair is the most important physical entity for quantum technologies such as quantum computing, quantum sensing, and quantum networking [3-7]. In general, entanglement between two photons or particles is not clearly defined due to the lack of a definite phase relationship [8]. According to general understanding of the quantum correlation of the HOM effect, the paired photons have been treated as identical and independent particles, i.e. indistinguishable photons on a BS [9-12]. However, the concept of indistinguishability originates in quantum superposition between probability amplitudes in two paths created by a BS or double slits, resulting in quantum superposition [13]. In this context, the phase relationship between two photons must be inevitably defined because the BS is an interferometric system [14]. As we know, however, quantum mechanics has been developed with quantum operators for the creation and annihilation of photons without the need for phase information [1-12]. Based on this textbook-level quantum theory based on the particle nature of a photon, the HOM effect is well explained even without definite phase information [8]. The fringe in a Mach-Zehnder interferometer can also be explained with the quantum operators [8].

    Recently, a new quantum mechanical interpretation for this quantum feature has been presented based on the wave nature of photons [15]. According to this theory, the two-photon intensity correlation on a BS requires a definite phase relationship between two interacting photons. Moreover, a quantum protocol of generating higher-order nonclassical features corresponding to photonic de Broglie waves [16] has also been developed, called coherence de Broglie waves (CBW) via MZI superposition [17]. Furthermore, a phase-basis quantization theory has been proposed to explain CBWs as well as the HOM effect to satisfy the indistinguishability [18,19]. According to this new interpretation based on the wave nature of photons [17-19], the quantum feature of the HOM effect is nothing but destructive quantum interference between the probability amplitudes of two interacting photons in an interferometric system. For this, the phase basis of the BS is equal to the phase difference between two interacting photons. As is well known in an MZI or Young's double-slit experiments, probability amplitudes in path choices by each photon must be indistinguishable [13,20,21]. In other words, the probability of finding a photon in each path must be the same. Thus, the meaning of indistinguishability originates not in the (phase) independent photons but in the equal probability amplitudes in a phase basis choice. Here, the physics of indistinguishability in the quantum feature of two-photon intensity correlation on a BS is investigated using the wave nature of photons. For this, quantized phase bases of a BS

are randomly applied to both input photons to clarify the quantum feature. As a result, the origin of the quantum feature remains in the indistinguishability between phase-basis choices, resulting in randomness in finding photons in each path.

**Results**

Figure 1 shows a BS-based two-photon interferometric system, where Fig. 1(a) is for the two-photon intensity correlation known as the Hong-Ou-Mandel (HOM) effect [2,9-12]. Figures 1(b) is a modified scheme of Fig. 1(a) for verification purposes of the present analysis. For this, a one-input-two-output BS system is separately analyzed (see the Supplementary Information), where the output photons are used as input photons for the BS2 in Fig. 1(b). For the analysis, first, a classical method of coherence optics is conducted as a reference. Then, a typical quantum approach follows and the results are compared with the classical ones. These two results show a distinct discrepancy and thus reveal the uniqueness of quantum mechanics on photon bunching. Finally, a proposed method with phase information is performed for the same scheme, and the results are compared with the previous ones to verify the origin of indistinguishability of photons for the quantum feature.

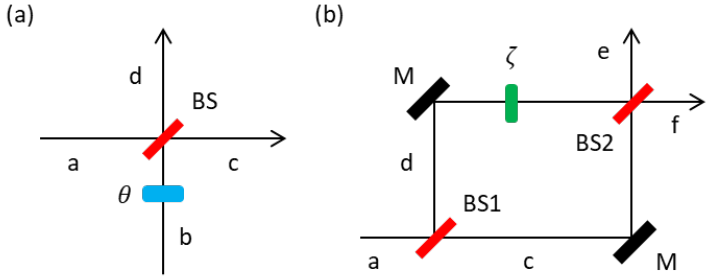

Fig. 1. Schematic of two-photon interactions on a beam splitter. (a) A HOM-type BS scheme. (b) A typical MZI scheme. The input photons originate from port 'a.' BS: a balanced (50/50) nonpolarizing beam splitter. θ, ζ: phase shifter. M: mirror.

In Fig. 1(a), two independent photons $E_j$ (j = a, b) are incident toward the BS from different ports, where the difference phase between two input photons is denoted by $e^{i\theta}$. Using coherence optics, the output photons are represented by [14]:

$$\begin{bmatrix} E_c \\ E_d \end{bmatrix} = \begin{bmatrix} 1 & \pm i \\ \pm i & 1 \end{bmatrix} \begin{bmatrix} E_a \\ E_b \end{bmatrix}$$
$$= E_0 \begin{bmatrix} 1 \pm e^{i\theta} \\ \pm i + e^{i\theta} \end{bmatrix}, \quad (1)$$

where $E_0$ is the amplitude of each photon. As a result, the intensities of output photons are $I_c = I_0(1 \mp sin\theta)$ and $I_d = I_0(1 \pm sin\theta)$. Remembering the classical definition of intensity in an ensemble average over θ, both output intensities become $\langle I_c \rangle = \langle I_c \rangle = I_0$ for the random phases of independent photons. The normalized mean coincidence measurement between two output photons is $\langle R_{cd} \rangle = cos^2(\theta)$, resulting in $\langle R_{cd} \rangle = 0.5$ for random phases. This is the classical lower bound in two-photon intensity correlation applied to classical particles [17,22]. This means that coherence optics should show various results depending on θ.

For a fixed θ $(= \pm \frac{\pi}{2})$, individual output intensities are still $\langle I_c \rangle = \langle I_c \rangle = I_0$ due to an equal probability between '+' and '−' in equation (1). However, the mean value of coincidence measurements is the average of individual coincident events, resulting in $\langle R_{cd} \rangle = 0$. This is the coherence quantum feature of photon bunching on a BS interpreted by pure coherence optics [15]. In other words, the difference in phase between two input photons on a BS plays a key role in the quantum feature generation as experimentally demonstrated, even though the authors did not mention the noclassical feature [23]. Thus, the quantum phenomenon of photon bunching belongs to coherence optics, but a measurement technique of coincidence detection reveals its hidden quantum nature [19].

Now, a quantum approach is applied to Fig. 1(a) using quantum operators of annihilation and creation processes in a BS [8]:

$$|1\rangle_a|1\rangle_b \rightarrow \frac{1}{2}(\hat{a}_c^\dagger + i\hat{a}_d^\dagger)(i\hat{a}_c^\dagger + \hat{a}_d^\dagger)|0\rangle_c|0\rangle_d$$
$$= \frac{i}{\sqrt{2}}(|2\rangle_c|0\rangle_d + |0\rangle_c|2\rangle_d). \quad (2)$$

Equation (2) represents the typical text-book understanding of quantum mechanical photon bunching on a BS. In equation (2), however, the photons resulting from creation and annihilation operators are presumed to have coherence via a two-photon coincidence process. In the same way, the presumption of random-phased photons indicates independent photons, like classical particles such as beads or stones. This is the critical discrepancy against coherence optics in the interpretation of quantum features in Fig. 1(a), where the BS is an interferometric with phase information. Assigning a mutual phase between paired photons in Fig. 1(a) does not violate quantum mechanics as with the EPR paradox [24] or violation of Bell's inequality with spin bases [25]. Here, the spin corresponds to the polarization or phase of a photon, resulting in the same category of basis relation in a Hilbert space. In general, assigning a relative phase to the two input photons is more appropriate to define correlation in an interferometric system (discussed below). In equation (2), it is assumed that a fixed phase relationship between two photons exists by the action of operators.

Regarding the proposed approach, the phase basis of a BS is newly defined as $\varphi_{BS}$ according to the wave nature of photons interacting with the BS in Fig. 1(a), resulting in either destructive or constructive interference: $\varphi_{BS} = \left\{-\frac{\pi}{2}, \frac{\pi}{2}\right\}$. The origin of the two phase bases can be found in the harmonic oscillation of a photon, $\boldsymbol{E}_j(r,t) = \boldsymbol{E}_0 e^{i(k\cdot r - \omega t)}$, where the phase modulus of the harmonic oscillation is $2\pi$. Thus, the destructive interference between two photons must be satisfied by a phase difference of $\pm\pi$. This is why there exists only two probability amplitudes in Born's rule for a photon interacting with a multi-slit system [26,27]. Thus, the phase bases of a BS must be related to the interacting photons with respect to the maximum number of interference fringes in a periodic modulus of $2\pi$ [18,28].

The general solution of the proposed method is obtained via the phase-basis superposition for two input photons. Regarding the basis superposition, four different cases are separately considered with an equal probability ratio. For the same phase-basis superposition for Fig. 1(a), the following representation is achieved:

$$\begin{bmatrix}E_c\\E_d\end{bmatrix} = \frac{1}{2}\left\{\begin{bmatrix}1 & \pm i\\ \pm i & 1\end{bmatrix} + \begin{bmatrix}1 & \pm i\\ \pm i & 1\end{bmatrix}\right\}\begin{bmatrix}E_a\\E_b\end{bmatrix}$$
$$= E_0\begin{bmatrix}1 & \pm i\\ \pm i & 1\end{bmatrix}\begin{bmatrix}1\\e^{i\theta}\end{bmatrix}. \quad (3)$$

Thus, $E_c = \frac{E_0}{2}(1 \pm ie^{i\theta})$ and $E_d = \frac{E_0}{2}(\pm i + e^{i\theta})$, where the corresponding intensities are $I_c = \frac{I_0}{2}(1 \mp \sin\theta)$ and $I_d = \frac{I_0}{2}(1 \pm \sin\theta)$, respectively. For random phased or independent photons, the mean value of each output photons in each output port is $\langle I_c\rangle = \langle I_c\rangle = I_0$ averaged over $\theta$. However, the normalized mean coincidence measurement between $I_c$ and $I_d$ is $\langle R_{cd}\rangle = \cos^2\theta$. For random $\theta$, $\langle R_{cd}\rangle = 0.5$ is obtained to satisfy the classical limit. For $\theta = \pm\frac{\pi}{2}$, however, $\langle R_{cd}\rangle = 0$ results, satisfying the same quantum feature of photon bunching as in equations (1) and (2). Thus, the present phase basis interpretation of equation (3) is verified for the same phase-basis superposition for two input photons under the definite phase relationship $\theta$.

For the opposite phase-basis superposition in Fig. 1(a), the following cases are separately considered:

(i) A symmetric case with opposite phase-basis superposition

$$\begin{bmatrix}E_c\\E_d\end{bmatrix} = \frac{1}{2}\left\{\begin{bmatrix}1 & i\\ i & 1\end{bmatrix} + \begin{bmatrix}1 & -i\\ -i & 1\end{bmatrix}\right\}\begin{bmatrix}E_a\\E_b\end{bmatrix}$$
$$= E_0\begin{bmatrix}1 & 0\\ 0 & 1\end{bmatrix}\begin{bmatrix}1\\e^{i\theta}\end{bmatrix}. \quad (4)$$

Thus, the resultant output intensities are $\langle I_c\rangle = \langle I_d\rangle = I_0$, regardless of $\theta$. The normalized mean value of the coincidence measurements is $R_{cd} = 1$, representing a coherence feature. Equation (4) violates photon bunching on a BS.

(ii) An antisymmetric case with opposite phase-basis superposition

$$\begin{bmatrix} E_c \\ E_d \end{bmatrix} = \frac{1}{2}\left\{ \begin{bmatrix} 1 & i \\ i & 1 \end{bmatrix} - \begin{bmatrix} 1 & -i \\ -i & 1 \end{bmatrix} \right\} \begin{bmatrix} E_a \\ E_b \end{bmatrix}$$
$$= -iE_0 \begin{bmatrix} 0 & 1 \\ 1 & 0 \end{bmatrix} \begin{bmatrix} 1 \\ e^{i\theta} \end{bmatrix}. \tag{5}$$

Thus, the results of each output intensity and the mean value of normalized coincidence measurements are the same as those in equation (4). These results with different phase-basis superposition never satisfy the quantum feature. To answer why, the validity of equations (4) and (5) is temporally held until Fig. 1(b) is analyzed for Fig. 1(a).

For Fig. 1(b), the following representation is obtained from coherence optics:

$$\begin{bmatrix} E_e \\ E_f \end{bmatrix} = \frac{E_0}{\sqrt{2}} \begin{bmatrix} 1 - e^{i\zeta} & i(1 + e^{i\zeta}) \\ i(1 + e^{i\zeta}) & -(1 - e^{i\zeta}) \end{bmatrix} \begin{bmatrix} 1 \\ 0 \end{bmatrix}, \tag{6}$$

where $\zeta$ is the MZI phase difference between two paths. For $\zeta = 0$, the output photon amplitudes are $E_e = 0$ and $E_f = i\sqrt{2}E_0$. For $\zeta = \pi$, however, the amplitudes of the output photons are reversed. Thus, the photon directionality or photon bunching in either output port of an MZI is satisfied for the coherence approach. Regarding the quantum approach with a single photon, the same MZI directionality has also been demonstrated, resulting in self-interference [20].

With the proposed phase-basis approach, the following two cases of representations are obtained for Fig. 1(b):

(iii)    For the symmetric case with the same phase-basis superposition

$$\begin{bmatrix} E_e \\ E_f \end{bmatrix} = \frac{1}{2}\left\{ \begin{bmatrix} 1 & \pm i \\ \pm i & 1 \end{bmatrix} + \begin{bmatrix} 1 & \pm i \\ \pm i & 1 \end{bmatrix} \right\} \begin{bmatrix} E_c \\ E_d \end{bmatrix},$$
$$= \begin{bmatrix} 1 & \pm i \\ \pm i & 1 \end{bmatrix} \begin{bmatrix} E_c \\ E_d \end{bmatrix}, \tag{7}$$

where $E_c = E_0/\sqrt{2}$ and $E_d = \pm i e^{i\zeta} E_0/\sqrt{2}$ from equation (1) for $E_b = 0$ because Fig. 1(b) is for the one input case (see the Supplementary Information). Thus, $E_e = 0$ and $E_f = \pm\sqrt{2}i e^{i\zeta} E_0$, where the corresponding intensities are $I_e = 0$ and $I_f = 2I_0$, satisfying the same quantum feature of photon bunching in Fig. 1(a) as well as the MZI directionality mentioned above. The antisymmetric case has no action due to the matrix cancellation.

(iv)    For the symmetric and antisymmetric case with opposite phase-basis superposition

$$\begin{bmatrix} E_e \\ E_f \end{bmatrix} = \frac{1}{2}\left\{ \begin{bmatrix} 1 & \pm i \\ \pm i & 1 \end{bmatrix} \pm \begin{bmatrix} 1 & \mp i \\ \mp i & 1 \end{bmatrix} \right\} \begin{bmatrix} E_c \\ E_d \end{bmatrix}$$
$$= \begin{bmatrix} 1 & 0 \\ 0 & 1 \end{bmatrix} \begin{bmatrix} E_c \\ E_d \end{bmatrix} \text{ or } i \begin{bmatrix} 0 & 1 \\ 1 & 0 \end{bmatrix} \begin{bmatrix} E_c \\ E_d \end{bmatrix}. \tag{8}$$

To calculate the $E_c$ and $E_d$ in equation (8), a one-input-two-output BS system is separately analyzed, resulting in $E_c = \sqrt{2}E_0$ and $E_d = 0$ for the '+' superposition case, but $E_c = 0$ and $E_d = i\sqrt{2}E_0$ for the '−' case. (see the Supplementary Information). Thus, the amplitudes of the output photons are $E_e = \sqrt{2}E_0$ and $E_f = i\sqrt{2}e^{i\zeta}E_0$ for the '+' case. For the '−' case, the amplitudes of the output photons are reversed, where $E_e = \mp\sqrt{2}e^{i\zeta}E_0$ and $E_f = i\sqrt{2}e^{i\zeta}E_0$. Thus, the corresponding intensities are $\langle I_e \rangle = \langle I_f \rangle = I_0$ and $\langle R_{ef} \rangle = 1$ in both cases, violating the MZI directionality as mentioned above. Thus, the opposite phase-basis superposition must be excluded from possible solutions. This is why equations (4) and (5) must be excluded. Thus, Fig. 1(a) is fully satisfied with the new approach of phase-basis superposition for the quantum features of photon bunching. In conclusion, the quantum feature on a BS is not satisfied for random phased photons but instead for phase correlated photons. In this manner, the disappearance of the wavelength-dependent interference fringe observed in ref. 23 is due to a wide bandwidth-causing coherence washout effects [29].

**Discussion**

From equations (1)-(3), the proposed interpretation based on the quantized phase bases of the photon-BS system satisfies the same quantum features of photon bunching as in the conventional interpretation of quantum mechanics based on the particle nature of photons. Remembering that indistinguishability originates in quantum superposition between two probability amplitudes according to Born's rule such as in Young's double-slit experiments and singe photon-based self-interference, the quantum feature on a BS cannot be accomplished by phase-independent photons. Instead, it is achieved by the phase-basis choices for the photons with equal probability, resulting in a quantum superposition between phase bases. As analyzed in equation (3), thus, the vague terms of indistinguishable photons in a conventional understanding of quantum mechanics is now clarified with the phase difference between photons via superposed amplitudes of the phase-basis choices. In other words, the two-photon intensity correlation requires not only a spatiotemporal overlap but also a definite relative phase between two photons.

**Conclusion**
The two-photon intensity correlation of the HOM effect on a BS was analyzed for the newly proposed method of phase-basis superposition of the photon-BS system. To verify the proposed method, both coherence and quantum approaches were also analyzed and compared with each other. As a result, the quantum feature of photon bunching on a BS was successfully demonstrated using the proposed phase-basis superposition of the photon-BS system. According to the proposed phase-basis superposition approach, the quantum feature needs a specific mutual phase relationship between two interacting photons on a BS. Thus, the conventional random phase-based quantum interpretation needs to be reconsidered, where mutual phase assignment does not violate quantum mechanics. In conclusion, the proposed method of phase-basis superposition views the quantum feature according to the wave nature of quantum mechanics and opens the door to deterministic quantum information processing. The key concept of indistinguishability in the two-photon quantum feature is not for independent photon characteristics but for the equal probability of phase basis choices, resulting in the same photon bunching via coincidence measurements.

**Acknowledgment**
This work was supported by the GIST-GRI 2021 and ICT R&D program of MSIT/IITP (2021-0-01810), Development of elemental technologies for ultra-secure quantum Internet.